\documentclass{elsart5p}
\usepackage{color,graphicx,amssymb}
\usepackage{epsfig,subfigure,changebar,picinpar,rotating}
\journal{Solid State Communications}
\begin{document}
\begin{frontmatter}
\title{Stochastic dynamics of microcavity polaritons}
\author{A.E. Pedraza\corauthref{cor}},
\corauth[cor]{Corresponding author.}
\ead{alv-pedr@uniandes.edu.co}
\author{L. Quiroga}
\address{Departamento de F\'{i}sica, Universidad de Los Andes, A.A. 4976, Bogot\'a D.C., Colombia}

\begin{abstract}
We study the time dependent polariton condensation as well as the
parametric scattering process of polaritons in a semiconductor
microcavity. Based upon a new stochastic scheme the dynamics for
both cases is fully analyzed. We show how the evolution of the
system is described by a set of stochastic differential
Schrodinger equations which in average reproduces the exact
dynamics. Furthermore, we underline the role that
Coulomb correlations plays in the polariton dynamics. Threshold
behaviors are well captured by the present approach. The results are in complete agreement with recent
experimental observations.
\end{abstract}
\end{frontmatter}

\section{Introduction}

Microcavity polaritons are quasiparticles created in the strong
coupling regime between photons in a microcavity and the excitonic
resonance of a semiconductor quantum well \cite{weiss}. Polaritons
satisfy Bose statistics (at least in the low density regime) and
are excellent candidates to study quantum effects at the
macroscopic level, e.g. Bose-Einstein Condensation (BEC) and
superfluidity. As a practical advantage that polariton systems
exhibit over other quasi-bosonic particles in condensed matter
systems is their easy experimental optical control
\cite{lang,SAV}. Particulary, polaritons can be optically injected
in the microcavity by external light impinging on the
semiconductor nanostructure, and their properties can be inferred
from the observation of the emitted light. The possibility of
obtaining a quantum fluid in a solid state system, with easy
control and integration potentialities, could also open a new
promising way to the implementation of quantum information
technologies.

The polariton condensation has being actively pursued, since no
BEC has being observed for a solid state system. Recently, a
possible quantum phase transition of polaritons was suggested
\cite{LAU}, and the spontaneous formation of a coherent polariton
state was reported in \cite{DEN}. This last result shows a
remarkable manifestation of the bosonic behavior of this short
living particles, but it has been interpreted as a polariton laser
phase transition. From the theoretical side, most existing works
\cite{LAU} pursue a strict analogy with the laser theory, where
the quantum fluctuations responsible for the phase transition
appear directly from the particle number conservation.

On the other hand, it is well known that due to their excitonic
component, polaritons are subjected to Coulomb interactions. This
nonlinearity is expected to trigger the quantum phase transition.
Moreover, the unexpected observation of a thermal-like intensity
far above the threshold \cite{DEN} does not match with the laser
picture, leaving the unanswered question of the role of
interactions, usually overlooked in the theory. A more recent
attempt to address this question has been made in \cite{SAR},
where the laser picture was left behind and polariton-polariton
interactions were taken into account. Nonetheless, this last study
is within the scope of a mean-field approximation (MFA), and thus
fails to give a proper description of the polariton dynamics above
the parametric threshold where high order Coulomb correlations are
more important than in the low density regime.

The present paper is devoted to gain more physical insight into
the polariton dynamics by analyzing two different dynamical
processes: (i) the time-dependent condensation and coherence
build-up of polaritons relaxing into the lowest branch states and
(ii) the time-dependent parametric scattering. Both processes will
be considered within the same theoretical approach based on the
stochastic unravelling of the dynamics of an interacting many-body
system embedded in a noisy environment. Partial realizations of
such a computational scheme have been previously considered. On
the one hand, {\it closed} quantum interacting systems have been
studied by stochastic unravellings in Ref.\cite{CARA} for bosons
and in Ref.\cite{jul} for fermions. On the other hand, simple {\it
open} quantum systems have been studied by following the
stochastic density matrix for the whole system plus environment in
Refs.\cite{BRE,LAC}. It is one of the main aims of the present
work to extend those partial developments to a complex many-body
open quantum system such as the one offered by microcavity
polaritons. In order to provide quantitative predictions, previous
stochastic schemes are extended to follow the evolution of a
polariton system coupled to a noisy environment where
interparticle interactions and environment effects can be treated
on the same footings.  This will allow us to transform the
evolution of an open interacting polariton system, into the
stochastic evolution of a single polariton \cite{CARA} separately
from the stochastic evolution of the environment \cite{BRE,LAC}.
As compared with similar approaches the advantages of this
stochastic scheme are manifold: first of all, the complete
dynamics, both transient and steady-state regimes, can be obtained
with full fluctuations due to interactions properly taken into
account. Second, a many-polariton problem can be mapped onto a
single polariton system driven by stochastic terms.

The paper is organized as follows: in Section \ref{stoc} a general
description of the stochastic scheme is presented. In Section
\ref{two} the polariton condensation dynamics with both Coulomb
interactions and phonon scattering is considered. We pay
particular attention to the quantum properties of the condensate
fraction and the second-order degree of coherence. The parametric
scattering of pump polaritons in the lower polariton branch and
the dynamics of formation of the eight-shaped final allowed states
is studied in Section \ref{scat}, where comparison to experimental
results \cite{lang} reveals the importance of including the
nonlinearities associated to polariton interactions, specially
above the threshold regime. Finally, in Section \ref{conc} we
sketch our concluding remarks.

\section{Stochastic scheme for interacting bosons}\label{stoc}

We start with a brief review of the main assumptions for a
stochastic scheme aiming to determine the evolution of a {\it
closed} many-boson system \cite{CARA}. The initial full
information of the system is encoded in the state vector
$\vert\psi\rangle=\vert N:\phi\rangle$ where N bosons are accommodated in the
same single particle state $\vert \phi \rangle$. Thus, one needs
only to find the appropriate evolution of $\vert \phi \rangle$ in
order to obtain the N-bosons system dynamics. For non interacting
particles this is of course trivial, but in the interacting case
the action of the Hamiltonian on the N-particle state vector
produces terms orthogonal to $\vert \phi \rangle$. This cannot be
taken into account by standard MFA, and hence the necessity to
introduce a noise term with specific properties. The first
non-trivial application of this method was performed in
Ref.\cite{CARL}.

For microcavity polariton systems at low temperatures, an always
present source of nonlinearity is the coupling with the acoustic
phonon reservoir. At first glance, the necessity to include this
reservoir, should leave the stochastic scheme just mentioned out
of the picture for polariton dynamics. However, as it has recently
been stated for simple open systems \cite{BRE,LAC}, the quantum
system and its environment can be described by a pair of decoupled
Stochastic Schrodinger Differential Equations (SSDE). The success
of this fictitious evolution is based on the adequate introduction
of stochastic terms. This will guarantee that when averaged over
the stochastic trajectories of both subsystems, the {\it exact}
whole dynamics is recovered. The central quantum system, however
decoupled from its surrounding in every individual trajectory,
will \emph{feel} the effect of the environment through a driven
noise term, allowing in this way to obtain the exact reduced
density matrix at any time. In the following of this Section we
will give a brief guideline on how the two just mentioned
stochastic schemes can be extended to consider polariton dynamics.

In general, the full Hamiltonian can be written as

\begin{eqnarray}\label{genham}
H_T=H_S+H_E+H_i
\end{eqnarray}
where $H_S$ describes the central quantum subsystem (including
interparticle interactions), $H_E$ is the Hamiltonian associated
to the environment and $H_i$ is the coupling Hamiltonian which we
assume can be expressed as $H_i=\sum_{\alpha}A_{\alpha}\otimes
B_{\alpha}$, where $A$ and $B$ are operators acting on the system
and environment, respectively. In order to obtain the formally exact density matrix
as the stochastic average over multiple realizations, each sample has to be taken as
the dyadics $\rho=\vert \Phi_1\rangle\langle \Phi_2 \vert$. Moreover, one of the main advantages of the present
approach is that the different kets and bras in the dyadics can be taken as the separable forms
$\vert\Phi_{\nu}\rangle=\vert\psi_{\nu}\rangle\otimes\vert\chi_{\nu}\rangle$ ($\nu=1,2$)
of system and environment state vectors, respectively. Of course, the stochastic average
reproduces the true entanglement between quantum system and environment, as it should be.
The evolution of the whole system under the Hamiltonian in Eq.(\ref{genham}) is then given by the set of SSDEs \cite{BRE}

\begin{eqnarray}\label{sys}
d\vert \psi_{\nu}(t)\rangle=-idtH_S\vert \psi_{\nu}\rangle-idt\sum_{\alpha}dW_{\alpha
\nu}(t)A_{\alpha}\vert \psi_{\nu}(t)\rangle
\end{eqnarray}

\begin{eqnarray}\label{env}
d\vert \chi_{\nu}(t)\rangle=-idtH_E\vert \chi_{\nu}\rangle+\sum_{\alpha}dW_{\alpha \nu}(t)B_{\alpha}\vert \chi_{\nu}(t)\rangle
\end{eqnarray}
with noise terms satisfying
$E[dW_{\alpha'\nu'}dW_{\alpha\nu}(t)]=\delta_{\alpha',\alpha}\delta_{\nu',\nu}$
and  $E[dW_{\alpha\nu}]=0$ ($E[...]$ means the average over
stochastic realizations). By averaging the dyadics built on the
states evolving as in Eq.(\ref{sys}) and Eq.(\ref{env}), the exact
Liouville equation for the whole density operator is found. This
has been extensively discussed in Ref.\cite{BRE} within a quantum
jump approach and alternatively in Ref.\cite{LAC} using a
diffusion method.

Assuming the Hamiltonian $H_S$ in Eq.(\ref{genham}) conserves the
number of particles, one can consider the dynamics of a closed
quantum system due to the separability from the environment.
Hence, the new idea is to describe the evolution of the N
interacting particles, now in an open situation, as occupying the
single particle state vector $\vert \phi\rangle$. In order to find
the stochastic evolution within this prescription, the new form of
the state vector $\vert \psi\rangle=\vert N:\phi\rangle$ has to be
adapted to Eq.(\ref{sys}), verifying that the SSDEs once again
yield to the formally exact dynamics. This procedure implies that

\begin{eqnarray}\label{ST-single}
d\vert \phi_{\nu}\rangle=-idtG_{\nu}\vert\phi_{\nu}\rangle+dM_{\nu}\vert\phi_{\nu}\rangle-i\sum_{\alpha}\tilde a_{\alpha}\vert \phi_{\nu}\rangle dW_{\alpha\nu}(t)
\end{eqnarray}
where $G_{\nu}$ contains inter-particle interaction effects
treated at the level of MFA while $dM_{\nu}$ is a noise term
responsible for capturing the particle-particle interaction
effects beyond MFA. Finally, the third term in
Eq.(\ref{ST-single}) carries the net effect of the environment
coupled to the system. The notation $\tilde a_{\alpha}$ is used to
represent a system's coupling operator projected into the single
particle space.

It is important to note that within this stochastic approach, where
system and environment are always \emph{separable} for each individual realization, the
calculation of any system's observable, after tracing over the
environment, is given by

\begin{eqnarray}\label{observable}
\langle \hat O(t) \rangle=E[\langle\psi_2(t)\vert
\hat O(t) \vert
\psi_1(t)\rangle\langle\chi_2\vert\chi_1\rangle].
\end{eqnarray}
With this in mind, a complex problem where bosonic particles
interact with each other and with their surroundings, has been
reduced to a set of SSDEs for a single particle,
Eq.(\ref{ST-single}), decoupled from its environment,
Eq.(\ref{env}). The price to be paid for this formidable reduction
of complexity is that a large number of stochastic realizations
has to be performed in order to recover the exact solution.
Obviously, the results obtained by this procedure are beyond MFA while
additionally particle-particle interactions as well as environment
effects (phonons) can be treated on equal grounds.

\section{Spontaneous coherence buildup of microcavity polaritons}\label{two}

\begin{figure}
\centering
\epsfig{file=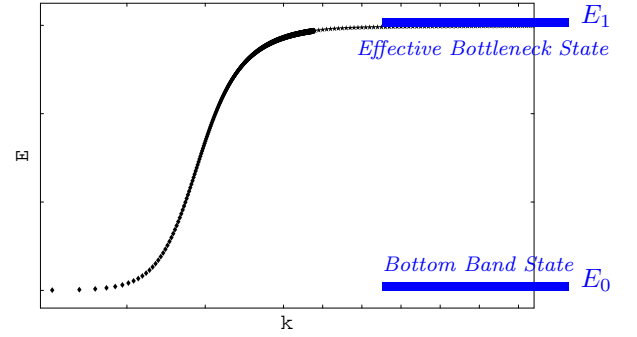,width=0.8\linewidth}
\put(10,117){\color{blue}\linethickness{3pt}\line(-1,0){70}}
\put(15,117){\color{blue} $E_1$}
\put(-70,105){\color{blue} \small \emph{Effective Bottleneck State}}
\put(10,17){\color{blue}\linethickness{3pt}\line(-1,0){70}}
\put(15,17){\color{blue} $E_0$}
\put(-60,23){\color{blue}\small \emph{Bottom Band State} }
\caption{Lower polariton energy branch versus total  momentum in a
linear-log scale. The excited state is represented by the
effective bottleneck state, which is at the edge of the flat
exciton-like dispersion region. The ground state is the zero
momentum state in the bottom of the dispersion
curve.}\label{disper}
\end{figure}

In a typical photoluminescence experiment under nonresonant
excitation, the strong energy dispersion of microcavity polaritons
leads to a bottleneck relaxation dynamics \cite{TAS}, where
polaritons go to the flat exciton-like region of the energy
dispersion curve. From this point on, polaritons relax into the
bottom states of the lower branch. Using these considerations a
simplified model of polariton dynamics is an effective two-level
system \cite{SAV}, where the ground state corresponds to ${\bf
k}=0$ which is non-degenerate ($E_0$), and the excited or
bottleneck set of states is macroscopically degenerate ($E_1$)
(see Figure \ref{disper}).

In the lower polartion branch the Hamiltonian describing both Coulomb and phonon scattering is

\begin{eqnarray}\label{pol-ham}
H&=&\sum_{k}\omega_k p^{\dagger}_k p_k + \sum_{q}\omega_q b^{\dagger}_q b_q + H_C+H_{ph}\\
H_C&=&\frac{1}{2}\sum{kk'q}v^{(q)}_{kk'}p^{\dagger}_{k+q}p^{\dagger}_{k'-q}p_{k'}p_k\\
H_{ph}&=&\sum_{kk'q}g^{(q)}_{kk'}(b^{\dagger}_q p^{\dagger}_k p_{k'}+H.C.)
\end{eqnarray}
where $p_k$ is the Bose annihilation operator for a polariton with
momentum $k$. The excitonic component of the polaritons is
responsible for Coulomb interactions, which is captured in $H_C$
with the coupling strength $v^{(q)}_{kk'}$. The semiconductor
environment is modelled by acoustical phonons distributed
according to the lattice temperature. The corresponding phonon
annihilation operator is $b_q$ with the usual Bose commutation
relations $[b_q,b^{\dagger}_{q'}]=\delta_{qq'}$. The
polariton-phonon scattering is responsible for decoherence and is
represented in Eq.(\ref{pol-ham}) as $H_{ph}$ with $g^{(q)}_{kk'}$
as the coupling strength. Note that this polariton Hamiltonian is
number conserving. The dynamical condensation of microcavity
polaritons will be studied using the stochastic formalism exposed
in the latter Section. The relevant material parameters
\cite{lang} are typical for a GaAs-based microcavity: heavy-hole
exciton Rabi splitting $\hbar\Omega_{R}=1.82meV$, the detunning
with the confined photon mode $-2.4meV$ and the effective index of
the cavity mode 3.5. The quantization area $A=100\mu m^2$
determines the Coulomb interaction \cite{CIU} and the phonon
coupling terms \cite{TAS}.

\begin{figure}
\centering
\epsfig{file=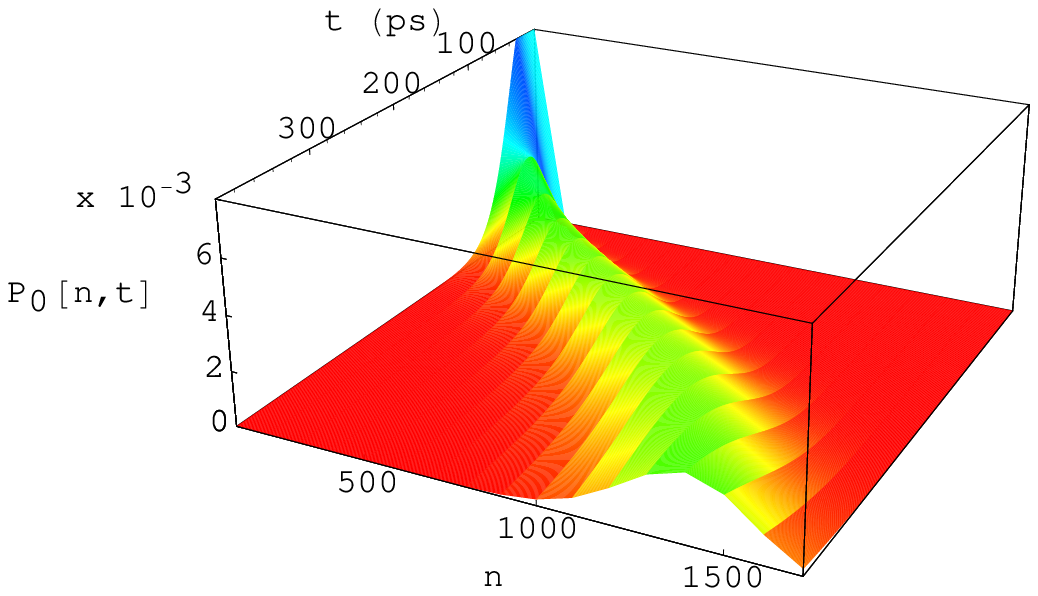,width=0.49\linewidth}
\epsfig{file=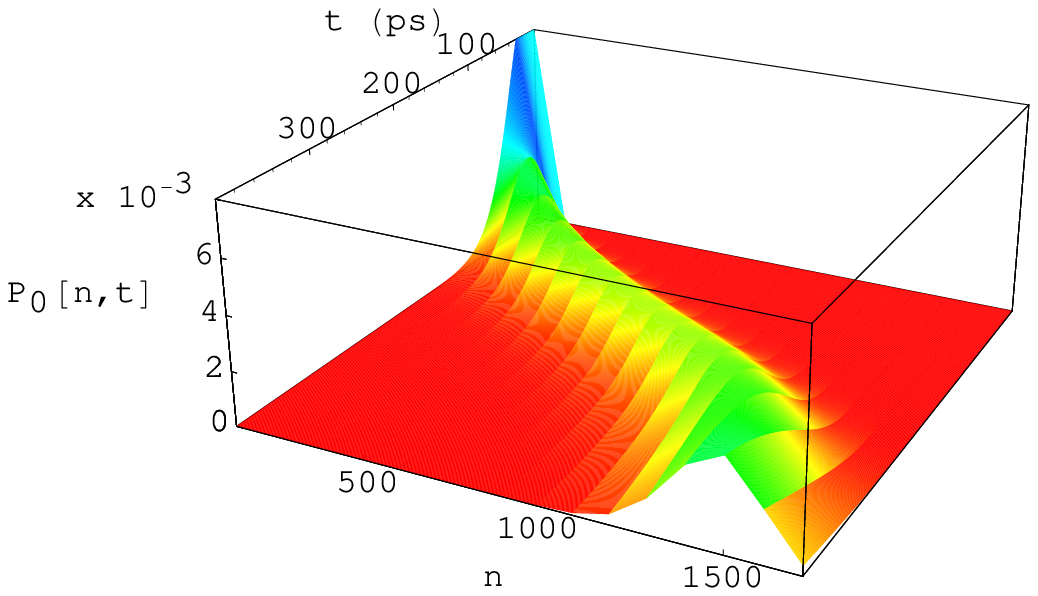,width=0.49\linewidth}
\put(-250,70){\tiny{(a)}}
\put(-120,70){\tiny{(b)}}
\\
\epsfig{file=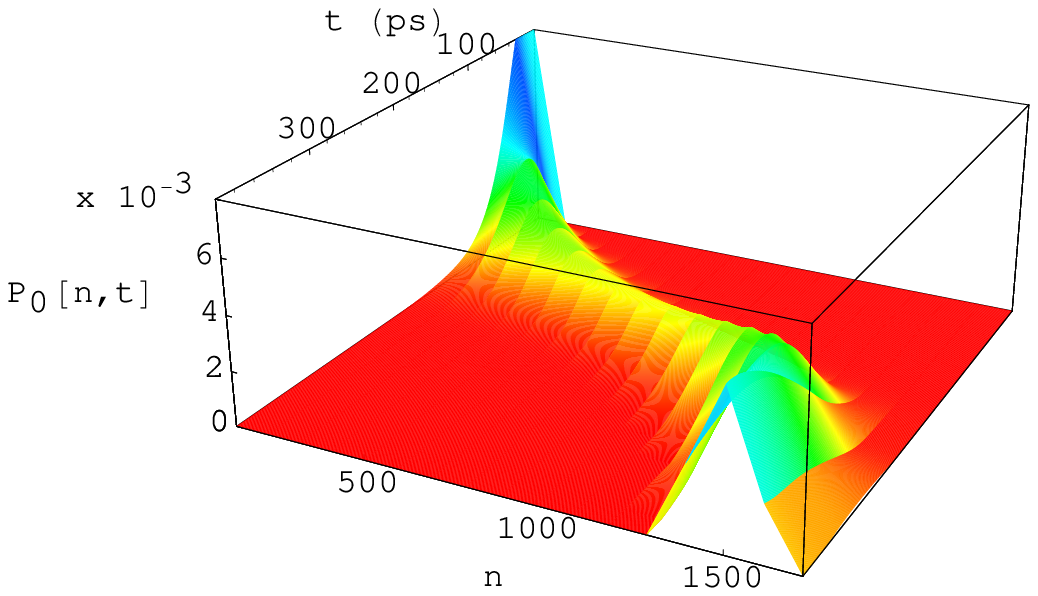,width=0.49\linewidth}
\epsfig{file=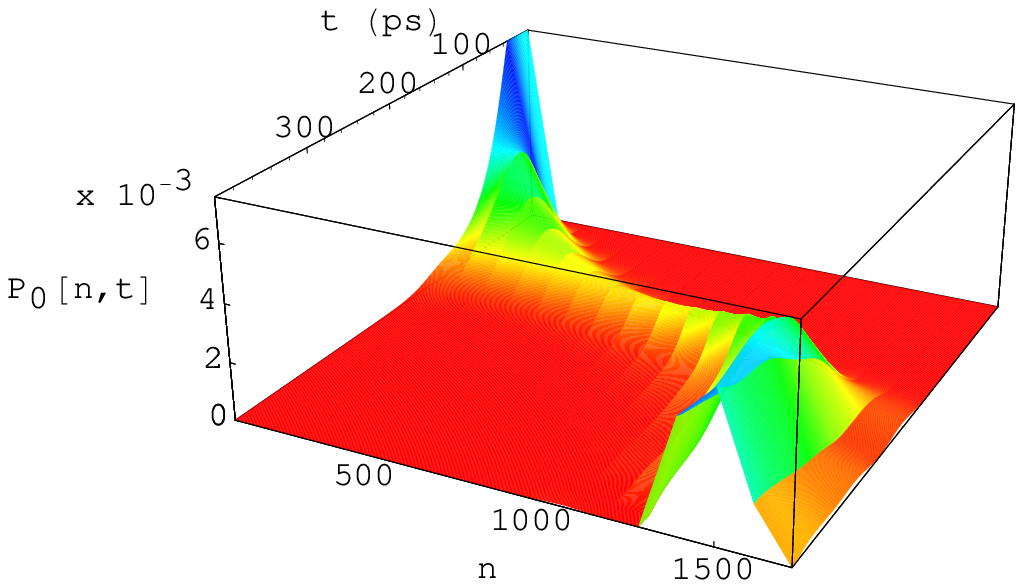,width=0.49\linewidth}
\put(-250,70){\tiny{(c)}}
\put(-120,70){\tiny{(d)}}
\caption{Ground state
probability distribution $P_0(n,t)$ for different polariton
interaction strengths: (a) $\gamma=0$, (b) $\gamma=0.4$, (c)
$\gamma=0.8$ and (d) $\gamma=1$, averaged over $5\times 10^5$ realizations. The total number of polaritons is
$N=1600$ and lattice temperature T=0.} \label{pro-dis}
\end{figure}

We consider N interacting polaritons in the presence of a
semiconductor environment at zero temperature. Since, the
effective two level model does not consider the relaxation from
high energy states into the effective bottleneck state, the
initial state consists of the total number of polaritons randomly
distributed in the excited energy level (importantly, no seed is
needed for level $E_0$). Indeed, a zero temperature environment
will force the polaritons into their ground state following a
non-trivial evolution due to polariton interactions. Our main
interest is to give a full description of the ground state
statistics. We first calculate the ground state probability
distribution $P_0(n,t)$. This is initially a delta function
centered at $n=0$, since all polaritons are initially in the
excited state. For comparison purposes, an overall factor,
$\gamma$, is introduced in the Coulomb interaction term in order
to account for the effective strength of two body Coulomb
interactions: $\gamma=0$ corresponds to no Coulomb interaction. As
is clearly shown in Figure \ref{pro-dis}, for a fixed total number
of polaritons $N=1600$, the polariton system decays monotonically
into its ground state when interactions are not included. As the
interactions are \emph{turned on}, the system's dynamics changes
completely. Polaritons start with a \emph{slow} relaxation
dynamics, but after a finite time the ground state becomes rapidly
populated with a macroscopic fraction of $N$.

\begin{figure}
\centering
\epsfig{file=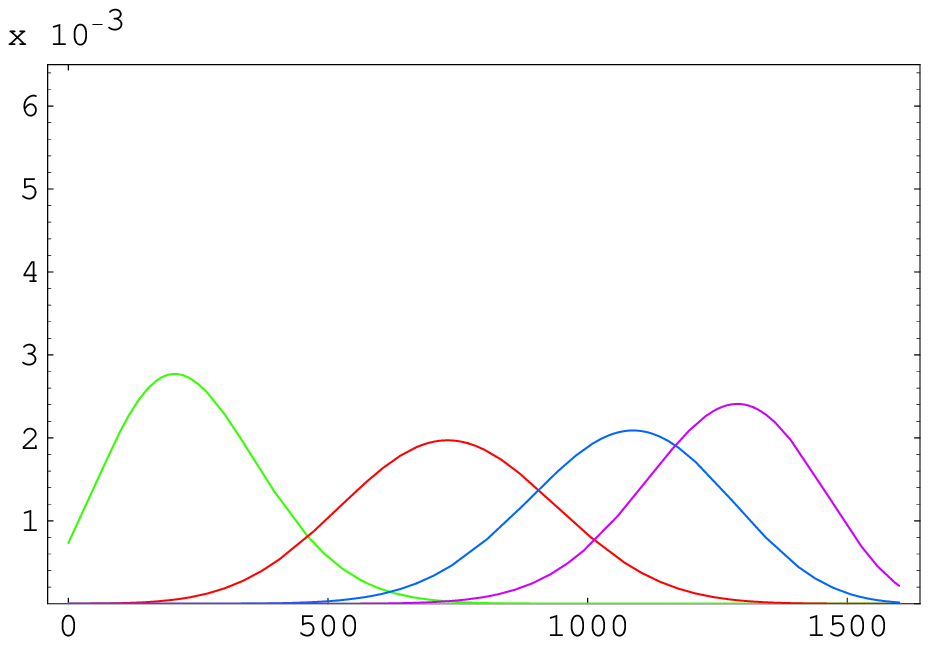,width=0.49\linewidth}
\epsfig{file=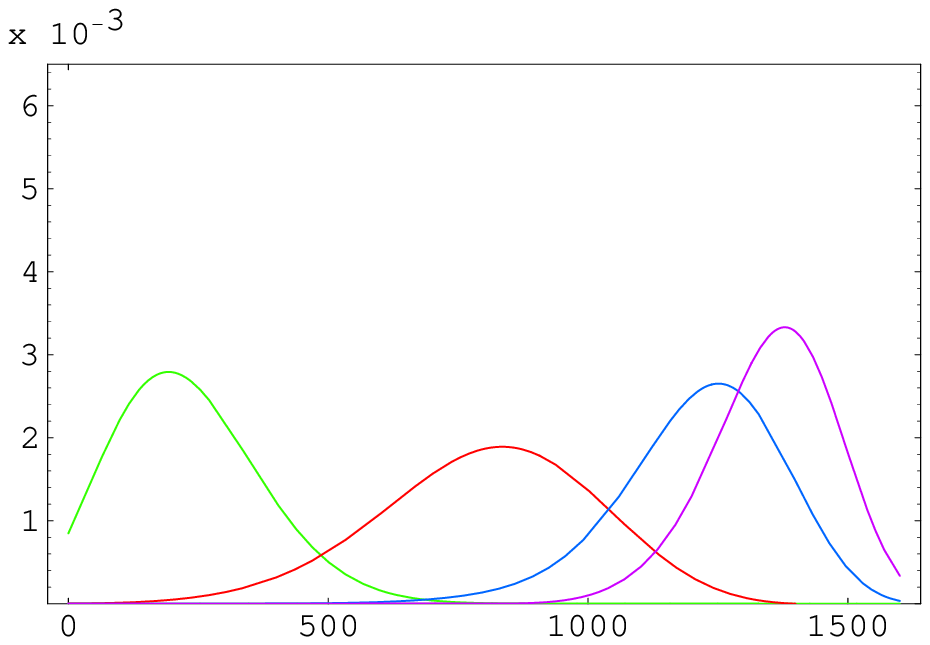,width=0.49\linewidth}
\put(-230,60){\tiny{(a)}}
\put(-195,60){\small $\gamma=0$}
\put(-105,60){\tiny{(b)}}
\put(-70,60){\small $\gamma=0.4$}
\put(-255,25){\rotatebox{90}{\small $P_0(n,t)$}}
\\
\epsfig{file=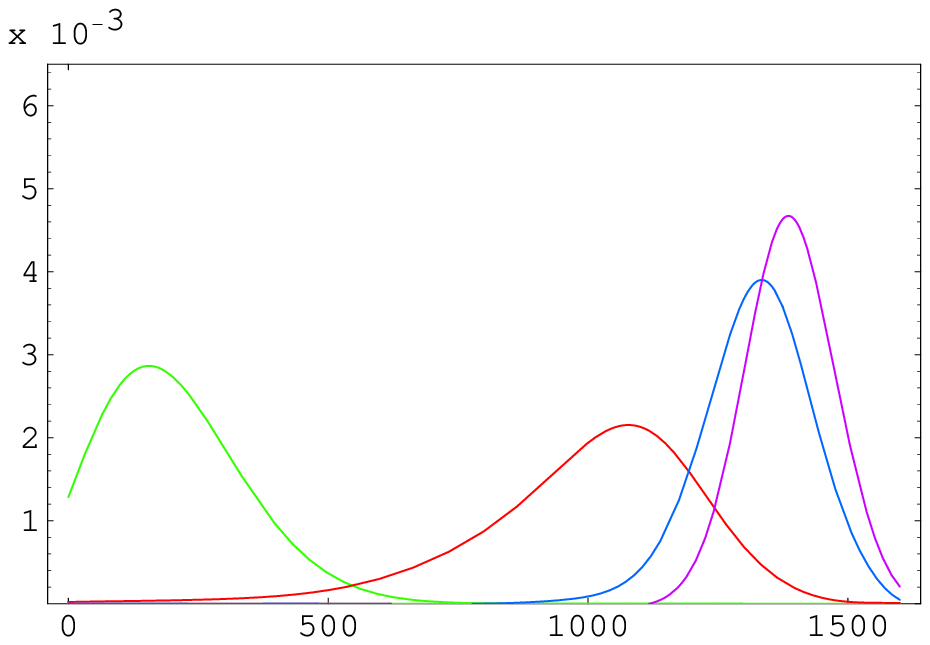,width=0.49\linewidth}
\epsfig{file=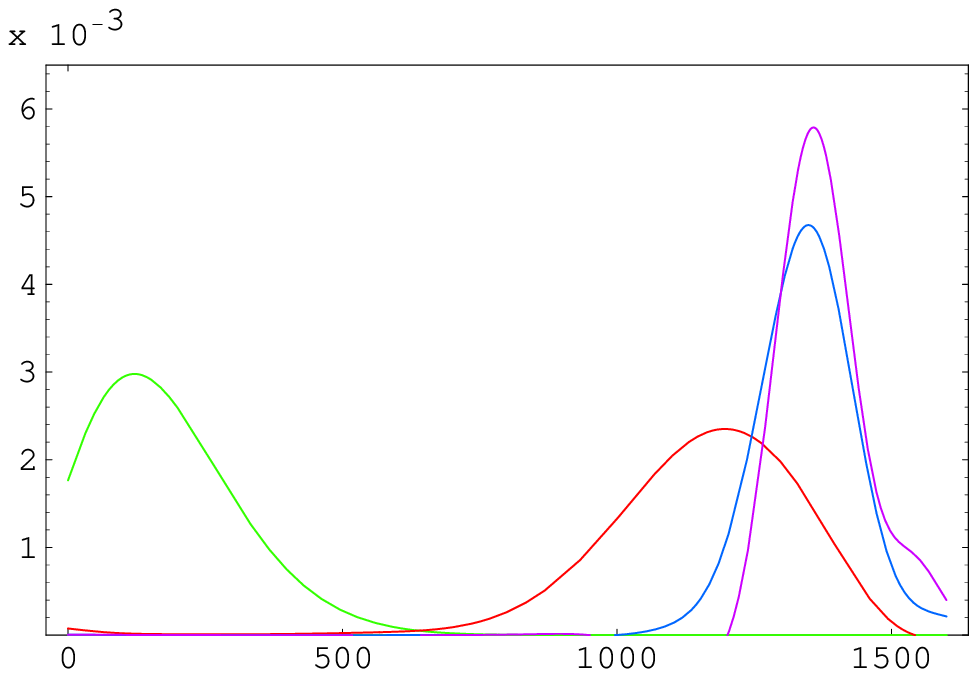,width=0.49\linewidth}
\put(-230,60){\tiny{(c)}}
\put(-195,60){\small $\gamma=0.8$}
\put(-105,60){\tiny{(d)}}
\put(-70,60){\small $\gamma=1$}
\put(-185,-5){n}
\put(-60,-5){n}
\put(-255,25){\rotatebox{90}{\small $P_0(n,t)$}}
\caption{Projections of the ground state probability distribution $P_0(n,t)$ at four various times: 100 ps (green), 200 ps (red), 300 ps (blue) and 400 ps (purple). Results are shown for N=1600, $5\times 10^5$ realizations and different polariton interaction strengths: (a) $\gamma=0$, (b) $\gamma=0.4$, (c) $\gamma=0.8$ and (d) $\gamma=1$.}
\label{distribution}
\end{figure}

The Gaussian-like distribution for vanishing inter-particle
interactions is evidently different from the Poisson-like
distribution when the interactions are fully included, as
displayed in Fig.\ref{distribution}. This latter feature is a
clear signature of an interaction driven coherence buildup process
for which an \emph{off-diagonal-long-range order} (ODLRO) can develop. It is well known \cite{PORRAS} that ODLRO does not implies a finite value of $\langle p_0\rangle$, in agreement with our results.

\begin{figure}
\centering
\epsfig{file=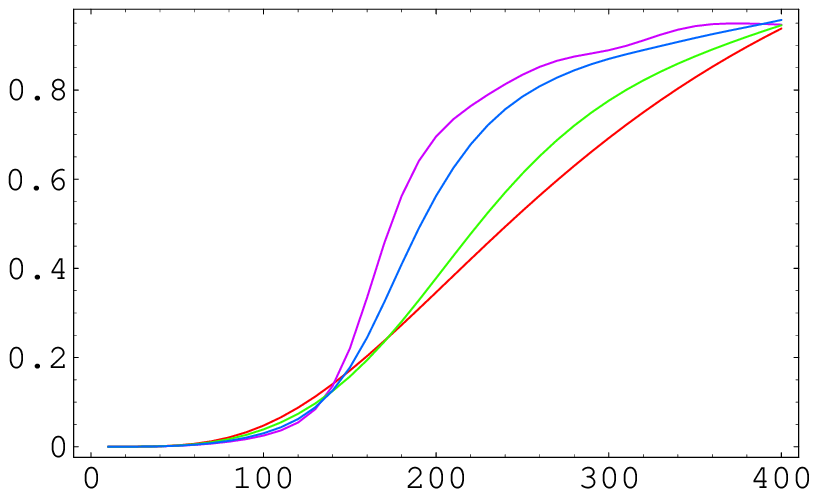,width=0.46\linewidth}
\epsfig{file=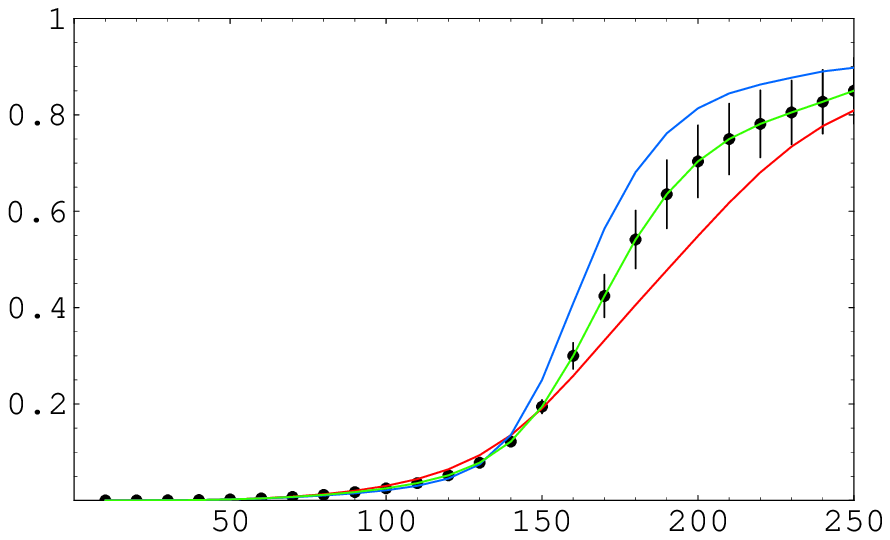,width=0.46\linewidth}
\put(-220,60){\tiny{(a)}}
\put(-100,60){\tiny{(b)}}
\put(-245,35){$\eta$}
\put(-180,-5){\small t (ps)}
\put(-70,-5){\small t (ps)}
\caption{Time evolution of the
coherence degree parameter $\eta$: (a) $N=1600$ and different
interaction strengths $\gamma=0$ (red), $\gamma=0.4$ (green),
$\gamma=0.8$ (blue) and $\gamma=1$ (purple). (b) Different number
of polaritons ($\gamma=1$) $N=$ 1400 (red), 1600 (green) and 1800
(blue). Lattice temperature T=0 and $5\times 10^5$ stochastic trajectories. Dispersion bars are shown for
the intermediate case of $N=1600$.} \label{coherence1}
\end{figure}

The so-called coherence degree parameter is defined as
$\eta=2-g^{(2)}(0)$, with $g^{(2)}(0)=\langle p^{\dagger 2}_0
p^{2}_0\rangle$. This parameter takes values between 0 and 1 for
states ranging from chaotic/thermal to coherent. Using this
parameter the full statistical properties of the ground state are
unravelled. Starting with an incoherent state (N polaritons
randomly distributed in the excited level), the coherence rises as
the ground state becomes populated. This is clearly shown in
Fig.\ref{coherence1}a. Again, the most interesting observation
about this evolution is the difference between the smooth growing
of coherence in the non-interacting regime as compared with the
sudden rise of coherence for the interacting case, complementing
the results depicted in Fig.\ref{distribution}. A threshold
behavior is clearly observed in Fig.\ref{coherence1}b. Once the
number of polaritons, or equivalently the density, is increased
beyond a certain value the coherence rises suddenly. A possible
phase transition triggered by Coulomb interactions is thus
reinforced. Moreover, since no high order correlations are in
principle discarded by the present method, as it is usually the
case for condensed matter systems \cite{SCH1,SCH2} (excitons) and
\cite{SAR} (polaritons), we can quantify their influence on the
polariton condensation dynamics.

Finally, we would like to remark some of the most important
aspects of these results. First of all, the threshold behavior
that governs the dynamics was already discussed in Ref.\cite{LAU},
and reported results are in good agreement with ours. This gives
us confidence that the stochastic scheme we used is indeed a very
practical tool for the study of complex systems like those found
in semiconductor microcavities. In particular, effects due to the
semiconductor environment are responsible for the relaxation of
the system, leading to a continuous accumulation of polaritons in
the same final state at low temperatures. Nevertheless,
correlations generated by interactions play a crucial role in the
polariton condensation dynamics. While a coherence degree
parameter close to 1 can only be interpreted as a criteria for
polariton lasing \cite{LAU}, instead of BEC in a
quasi-two-dimensional nanostructure, signs of a true phase
transition are present in our results as a consequence not only of
the bosonic behavior but from quantum fluctuations provided by
polariton interactions. Since the appearance of the phase
transition is governed by Coulomb interactions, our results point
toward a BEC type behavior rather than a polariton laser.

\section{Polariton parametric scattering}\label{scat}

In recent experiments \cite{lang,SAV} polaritons are optically
excited at a well defined in-plane momentum {\bf k}, allowing a
direct control of their dynamics. The strong Coulomb interaction
results in a spontaneous parametric scattering of polariton pairs
(signal and idler) according to momentum and energy conservation
criteria. For resonant optical excitation, the scattering leads
into a final state 8-shaped distribution in {\bf k} space, as
first suggested by Ciuti \emph{et al} \cite{CIU} and
experimentally confirmed in Ref.\cite{lang}. We use the present
stochastic scheme to study the dynamics of initially pumped
polaritons in a particular momentum state {\bf k} in the lower
polariton branch. Once again, the advantages of the present
approach are manifold. Similar to the condensation of polaritons
studied in the previous Section, the theoretical discussion of
parametric scattering has been based upon MFA, and thus the
results are only valid below the threshold of parametric
luminescence. Thus, the study of the macroscopic behavior of
interacting polaritons in a relaxing environment has been poorly
investigated \cite{CAR}. Moreover, the dynamical behavior of
condensed matter quasi-particles in an out of equilibrium regime
can be systematically studied within our approach. In this Section we abandon the effective
two-level approximation and resort to an extended multi-level description.

\begin{figure}
\centering \epsfig{file=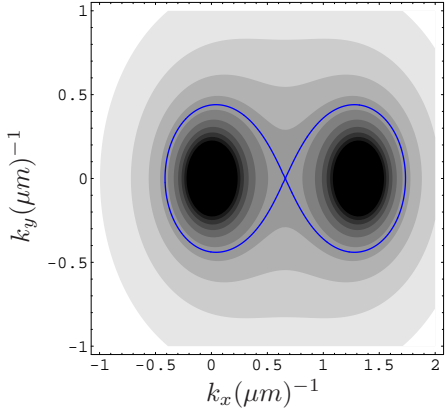,width=0.6\linewidth}
\put(-90,-5){$k_x (\mu m)^{-1}$} \put(-165,55){\rotatebox{90}{$k_y
(\mu m)^{-1}$}} \caption{Contour plot of the energy balance
equation $\vert E_{k_s}+E_{k_i}-2E_{k_p}\vert$. The blue line
represents the set of final states for two pump polaritons at
${\bf k_p}=\{0.66,0\}\mu m^{-1}$ spontaneously scattered
conserving both energy and momentum.} \label{eight}
\end{figure}

\begin{figure}
\centering
\epsfig{file=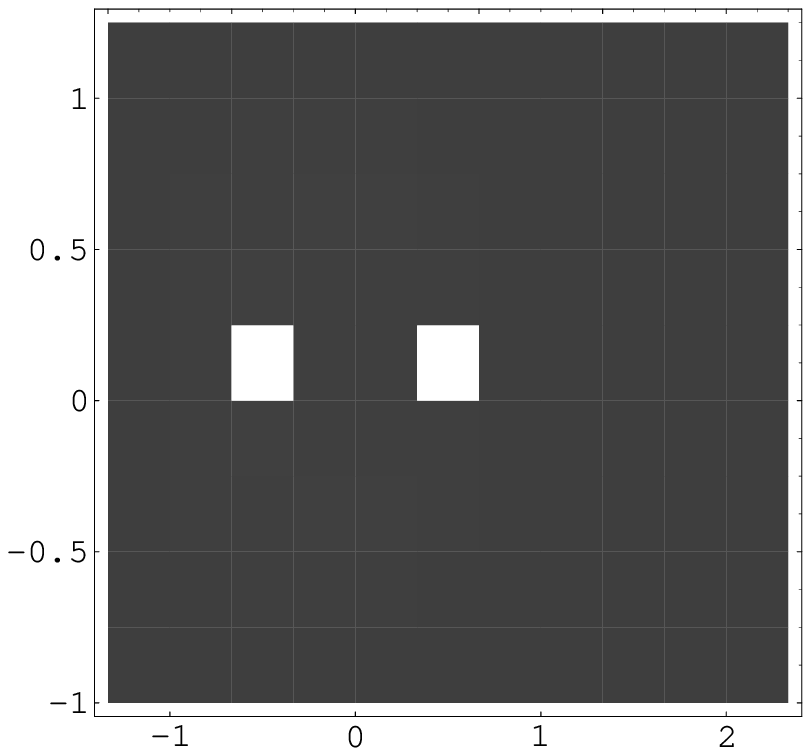,width=0.48\linewidth}
\epsfig{file=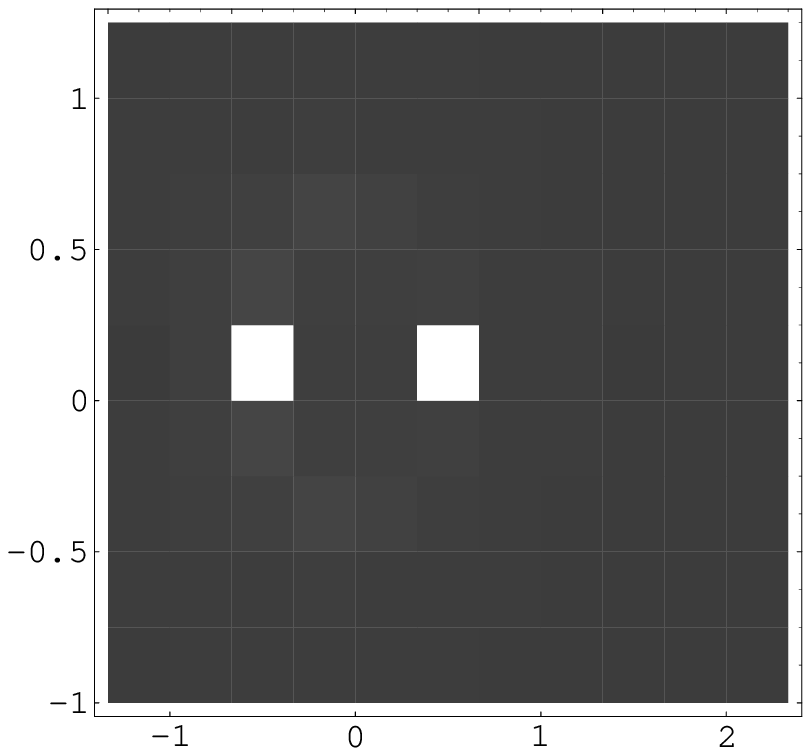,width=0.48\linewidth}
\put(-222,105){\textcolor{white}{\tiny{(a)}}}
\put(-100,105){\textcolor{white}{\tiny{(b)}}}
\put(-255,40){\rotatebox{90}{$k_y (\mu m)^{-1}$}}\\
\epsfig{file=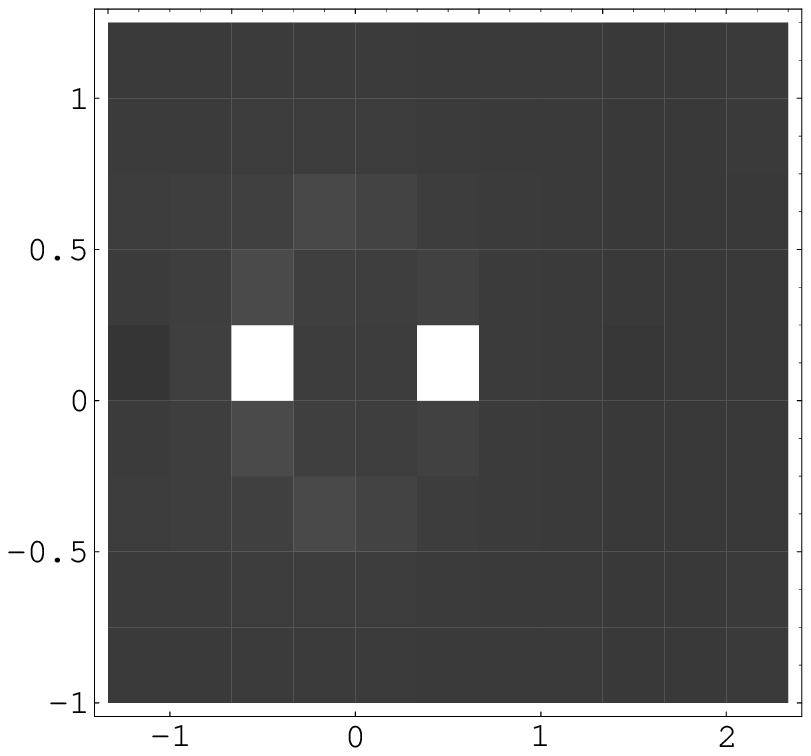,width=0.48\linewidth}
\epsfig{file=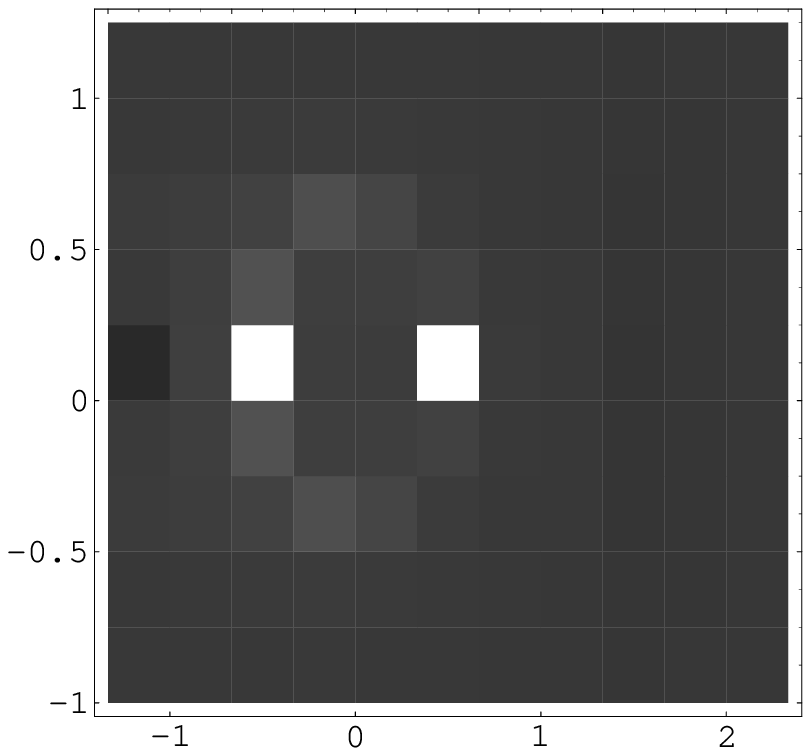,width=0.48\linewidth}
\put(-222,105){\textcolor{white}{\tiny{(c)}}}
\put(-100,105){\textcolor{white}{\tiny{(d)}}}
\put(-200,-5){$k_x (\mu m)^{-1}$}
\put(-80,-5){$k_x (\mu m)^{-1}$}
\put(-255,40){\rotatebox{90}{$k_y (\mu m)^{-1}$}}\\
\caption{Population evolution in momentum space of N=1000
polaritons initially pumped at ${\bf k}=\{0.66,0\}\mu m^{-1}$. The
plots correspond to different times: (a) 20 ps, (b) 40 ps, (c) 60 ps (d) 80 ps, and $10^4$ simulations.} \label{N1000}
\end{figure}

For the numerical simulations that follows, the total Hamiltonian
is still given by Eq.\ref{pol-ham}, where the system's parameters
are the same as in Section \ref{two}. The possible final states
for the scattering of two polaritons pumped at ${\bf
k}=\{k_p,0\}$, into a signal and idler pair, are displayed in
Fig.\ref{eight}, where the long-time limit, in which both energy
and lineal momentum have to be conserved, is depicted as the
continuous line. This information is crucial for understanding the
many-polariton case that we discuss from now on. The initial state
is taken as N polaritons pumped at ${\bf k_p}=\{0.66,0\}\mu
m^{-1}$. The system is let to evolve and population in ${\bf k}$
space is monitored at different times. In Fig.\ref{N1000} contour
plots of the polariton population distribution are shown, for
N=1000 particles initially pumped. Clearly, a strong asymmetry in
signal and idler emission patterns is seen. In particular,
although there is a small population in the idler modes, it is not
comparable to the signal one.

%$E_k+E_{k_{idler}}-2E_{k_p}$,${\bf k_p}=\{0.66,0\}\mu m^{-1}$

%\begin{figure}
%\centering
%\includegraphics[width=8cm,height=8cm]{N1000.eps}
%\put(-205,210){\textcolor{white}{\tiny{(a)}}}
%\put(-100,210){\textcolor{white}{\tiny{(b)}}}
%\put(-205,100){\textcolor{white}{\tiny{(c)}}}
%\put(-100,100){\textcolor{white}{\tiny{(d)}}}
%\put(-235,40){\rotatebox{90}{$k_y (\mu m)^{-1}$}}
%\put(-235,150){\rotatebox{90}{$k_y (\mu m)^{-1}$}}
%\put(-180,-10){$k_x(\mu m)^{-1}$} 
%\put(-70,-10){$k_x (\mu m)^{-1}$}
%\caption{Population evolution in momentum space of N=1000
%polaritons initially pumped at ${\bf k}=\{0.66,0\}\mu m^{-1}$. The
%plots correspond to different times: (a) 20 ps, (b) 40 ps, (c) 60 ps (d) 80 ps, and $10^4$ simulations.} \label{N1000}
%\end{figure}

\begin{figure}
\centering
\epsfig{file=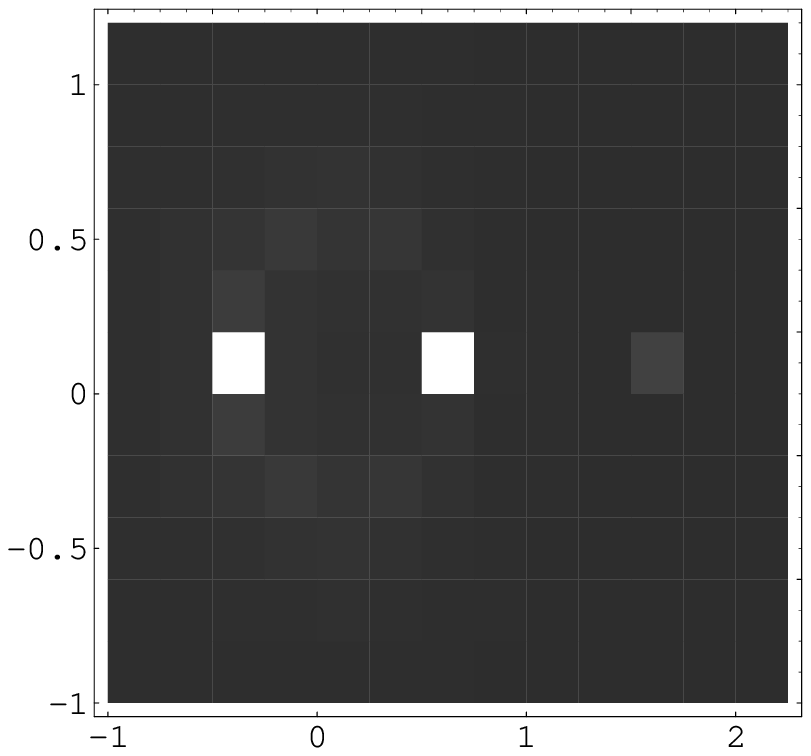,width=0.48\linewidth}
\epsfig{file=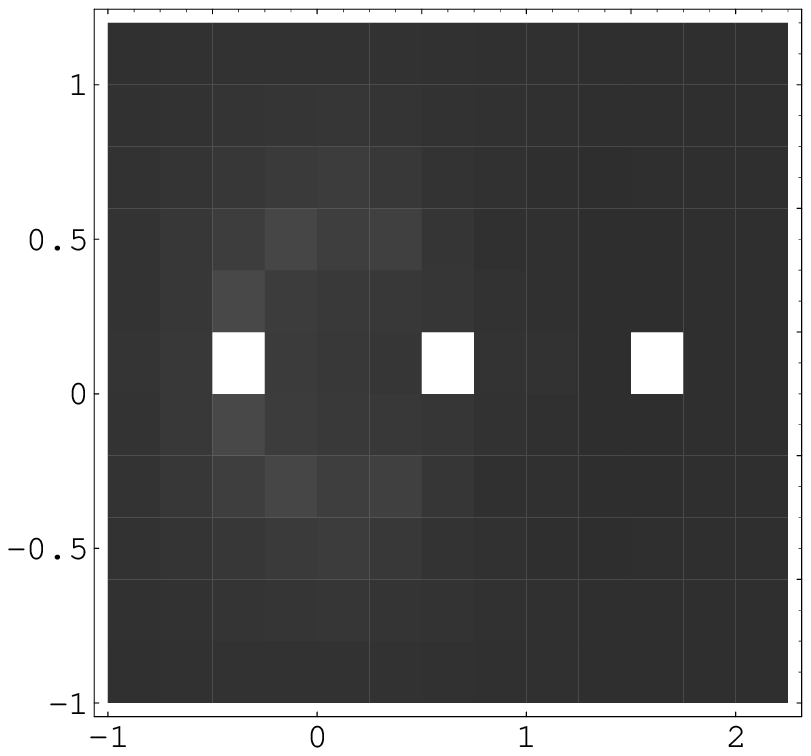,width=0.48\linewidth}
\put(-222,105){\textcolor{white}{\tiny{(a)}}}
\put(-100,105){\textcolor{white}{\tiny{(b)}}}
\put(-255,40){\rotatebox{90}{$k_y (\mu m)^{-1}$}}\\
\epsfig{file=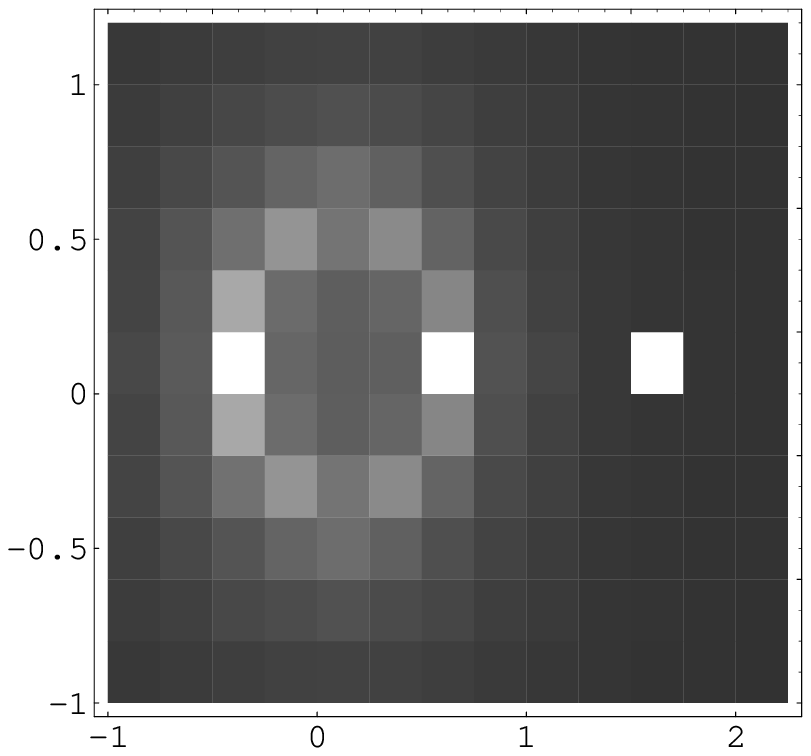,width=0.48\linewidth}
\epsfig{file=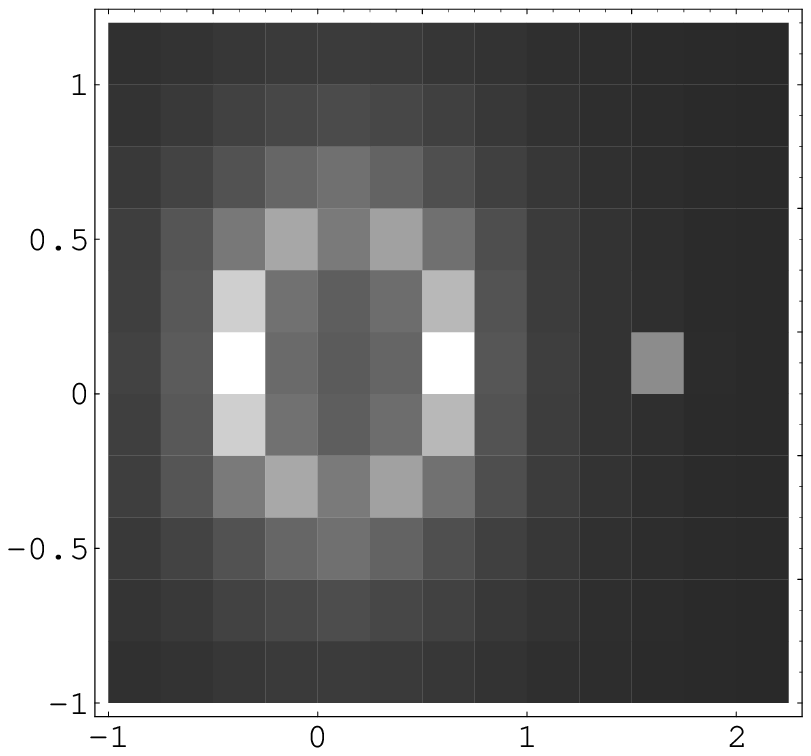,width=0.48\linewidth}
\put(-222,105){\textcolor{white}{\tiny{(c)}}}
\put(-100,105){\textcolor{white}{\tiny{(d)}}}
\put(-200,-5){$k_x (\mu m)^{-1}$}
\put(-80,-5){$k_x (\mu m)^{-1}$}
\put(-255,40){\rotatebox{90}{$k_y (\mu m)^{-1}$}}\\
\caption{Population evolution in the momentum space of N=5000
polaritons initially pumped at  ${\bf k}=\{0.66,0\}\mu m^{-1}$.
The plots correspond to different times (a) 20 ps, (b) 40 ps, (c)
60 ps, (d) 80 ps, and $10^4$ simulations.} \label{N5000}
\end{figure}

Above the parametric threshold a remarkable different behavior is
displayed. As the density of polaritons increases a more
significant fraction of the population is concentrated in lowest
momentum states, and thus an important intensity is registered in
the idler state opposed to the signal one, as can be seen in
Fig.\ref{N5000}. Similar results were reported in Ref.\cite{CAR}.
Furthermore, the finite intensity that rises in the idler state is
asyncronous with the evolution of the signal population, in close
agreement with experimental results in Ref.\cite{lang}. The
asymmetry in the shape and time of the formation dynamics of the
final state distribution in ${\bf k}$ space is a direct result of
the different multiple scattering processes affecting signal and
idler polaritons. As is well known, for higher momentum states
(idler modes), polaritons have a big excitonic component, which
implies a stronger coupling with the lattice phonons, and as a
consequence a fast decay in idler population. Even though in the
time range of the reported dynamics the dominant scattering
processes correspond to Coulomb interactions, some evidence of
decoherence in the idler states, due to the semiconductor
environment (phonons), is visible in Fig. \ref{N5000}c and
\ref{N5000}d.

%\begin{figure}
%\centering
%\includegraphics[width=8cm,height=8cm]{N5000.eps}
%\put(-205,210){\textcolor{white}{\tiny{(a)}}}
%\put(-100,210){\textcolor{white}{\tiny{(b)}}}
%\put(-205,100){\textcolor{white}{\tiny{(c)}}}
%\put(-100,100){\textcolor{white}{\tiny{(d)}}}
%\put(-235,40){\rotatebox{90}{$k_y (\mu m)^{-1}$}}
%\put(-235,150){\rotatebox{90}{$k_y (\mu m)^{-1}$}}
%\put(-180,-10){$k_x(\mu m)^{-1}$} 
%\put(-70,-10){$k_x (\mu m)^{-1}$}
%\caption{Population evolution in the momentum space of N=5000 
%polaritons initially pumped at  ${\bf k}=\{0.66,0\}\mu m^{-1}$.
%The plots correspond to different times (a) 20 ps, (b) 40 ps, (c)
%60 ps, (d) 80 ps, and $10^4$ simulations.} \label{N5000}
%\end{figure}

\begin{figure}
\centering \epsfig{file=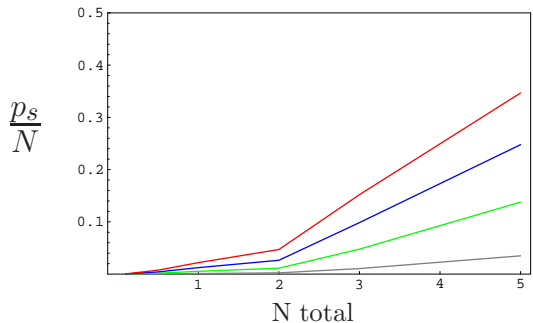,width=0.7\linewidth}
\put(-100,-10){N total} \put(-200,60){\Large $\frac{p_s}{N}$}
\caption{Population fraction of the signal mode ${\bf
k_s}=\{-0.5,0\}\mu m^{-1}$ as a function of the initially pumped
polaritons $N$ at different times: $t=5$ ps (gray), $t=10$ ps
(green), $t=15$ ps (blue) and $t=20$ ps (red). $10^4$ realizations
for each initial number of total polaritons.} \label{thres}
\end{figure}

In contrast to MFA, the threshold behavior is well captured by our
stochastic method. As the polariton density increases, a
dramatic change is observed in the signal population. While below
the threshold the signal population is weak, a sudden increase
close to the threshold is observed. In Fig.\ref{thres}, the signal
population is displayed for different times as a function of the
number of initially pumped polaritons. Close to $N=2000$, a clear change of
slope is registered for any time. The behavior of the idler mode populations (not
shown) is analogous.

\section{Conclusions}\label{conc}

In the present paper we have presented a stochastic study on the
dynamical properties of microcavity polaritons. In particular, we
reviewed the dynamical condensation of polaritons emphasizing on
the role of Coulomb interactions. Additionally, the threshold
behavior of the spontaneous parametric downconversion of
polaritons was also discussed. The numerical calculations have
been performed by means of stochastic evolution of a reduced state
vector in the single particle Hilbert space, separately from the
stochastic environment evolution. The resulting stochastic
differential equations allows reasonable reduction of the problem
which by direct numerical integration of the density matrix would
have being cumbersome. For this matter, the stochastic technique
turns into a very practical tool for many body problems in
condensed matter scenarios. Its validity was tested and compared positively
with previous theoretical and experimental results.

In the dynamical condensation of polaritons, we found that while
the semiconductor environment induces a slow relaxation process,
polariton-polariton interactions trigger the sudden development of
a macroscopic coherent state. This differs clearly from the
polariton laser picture, and other recent proposals which claim
that this behavior is just a consequence of the particle number
conservation \cite{LAU}. While a more detailed study on this
phenomenon is needed, our results favor the BEC type phase
transition of microcavity polaritons.

On the other hand, the dynamical process of parametric emission
was studied.  There exists an asymmetry in shape and time
dependence of the far field emission of signal and idler modes
governed by this kind of scattering processes. The threshold
behavior is also well captured by our stochastic method. This
corresponds to the sudden rise in signal and idler populations.
Coherent properties of this parametric emission process may also
be studied by the present method. Additionally, a straightforward possibility is to extend the present
stochastic method to finite lattice temperature situations.

This work was supported by COLCIENCIAS-1204-05-13614 (Colombia)
and Facultad de Ciencias-UniAndes.

\end{document}